**Title:** Comparing Theories for the Maintenance of Late LTP and Long-Term Memory: Computational Analysis of the Roles of Kinase Feedback Pathways and Synaptic Reactivation

12,390 words, 5 figures

**Authors:** Paul Smolen[1,*], Douglas A. Baxter[1,2], John H. Byrne[1]

**Affiliations:**

[1] Department of Neurobiology and Anatomy

W.M. Keck Center for the Neurobiology of Learning and Memory

McGovern Medical School at the University of Texas Health Science Center at Houston

6431 Fannin Street, Suite MSB 7.046, Houston, TX 77030

[2] Engineering and Medicine

Texas A&M Health Science Center

2121 Holcombe Blvd., Houston, TX 77030

**Correspondence address:**

Paul D. Smolen, Ph.D., Department of Neurobiology and Anatomy, W.M. Keck Center for the Neurobiology of Learning and Memory, McGovern Medical School of the University of Texas Health Science Center at Houston, Houston, TX 77030. Tel: 713-500-5579, Fax: 713-500-0623, E-mail: <u>Paul.D.Smolen@uth.tmc.edu.</u>

## Abstract

A fundamental problem in neuroscience is how memories are maintained from days to a lifetime, given turnover of proteins that underlie expression of long-term synaptic potentiation (LTP) or 'tag' synapses as eligible for LTP. One likely solution relies on synaptic positive feedback loops, prominently including persistent activation of $Ca^{2+}$/calmodulin kinase II (CaMKII) and self-activated synthesis of protein kinase M ζ (PKMζ). Recent studies also suggest positive feedback based on recurrent synaptic reactivation within neuron assemblies, or engrams, is necessary to maintain memories. The relative importance of these feedback mechanisms is controversial. To explore the likelihood that each mechanism is necessary or sufficient to maintain memory, we simulated maintenance of LTP with a simplified model incorporating persistent kinase activation, synaptic tagging, and preferential reactivation of strong synapses, and analyzed implications of recent data. We simulated three model variants, each maintaining LTP with one feedback loop: autonomous, self-activated PKMζ synthesis (model variant I); self-activated CamKII (model variant II); and recurrent reactivation of strengthened synapses (model variant III). Variant I requires and predicts that, for successful maintenance, PKMζ must contribute to synaptic tagging. Variant II maintains LTP and suggests persistent CaMKII activation could maintain PKMζ activity, a feedforward interaction not previously considered. However we note data challenging this feedback loop. In Variant III synaptic reactivation drives, and thus predicts, recurrent or persistent activity elevations of CamKII and other necessary kinases, plausibly contributing to empirically persistent elevation of PKMζ levels. Reactivation is thus predicted to sustain recurrent rounds of synaptic tagging and incorporation of plasticity-related proteins. We also suggest (model variant IV) that synaptic reactivation and autonomous kinase activation could synergistically maintain LTP. We propose experiments that could discriminate these maintenance mechanisms.

## Introduction

Long-term memory (LTM) lasting days or more necessitates synaptic structural changes and strengthening, and *de novo* protein synthesis (Davis and Squire 1984; Kandel 2001; Mayford et al. 2012). Long-term potentiation (LTP), and the invertebrate analog denoted long-term synaptic facilitation (LTF), are likely necessary for many types of memory formation (Asok et al. 2019; Lynch et al. 2004; Mirisis et al. 2016; Nabavi et al. 2014). Late LTP (L-LTP, lasting > ~2 h) requires protein synthesis (Abraham and Williams 2008; Frey et al. 1998; Stanton and Sarvey 1984) and can last at least a year (Abraham et al. 2002). Thus a fundamental concern is to determine mechanisms that preserve LTM for months or even a lifespan (Crick 1984; Holliday 1999; Roberson and Sweatt 1999; Schwartz 1993; Vanyushin et al. 1974).

One hypothesized mechanism involves persistent upregulation of local translation, or enhanced incorporation, of synaptic proteins commonly termed plasticity-related proteins (PRPs; Fig. 1) (Fonkeu et al. 2019). Increased translation or incorporation would be based on positive feedback loops specific to those synapses that have been strengthened to participate in LTM storage (Smolen et al. 2019). As illustrated in models, a common element of such positive feedback is a bistable switch in which the strength or weight of specific synapses, and biophysical variables such as kinase activities, switch post-stimulus from basal to persistently increased values (Bhalla and Iyengar 1999; Lisman 1985; Song et al. 2007; Zhang et al. 2016). Although synapse-specific mechanisms for maintenance of LTM may be supplemented by cell-wide mechanisms, such as upregulation of a specific set of genes (e.g., Kyrke-Smith and Williams, 2018) or neuronal excitability (Mozzachiodi and Byrne 2010), here we explore the necessity, sufficiency, and interactions among proposed synapse-specific positive feedback loops to maintain LTP. Synapse-specific feedback loops are necessary. In their absence, given protein turnover time scales of hours to days (Cohen et al. 2013; Eden et al. 2011), homeostatic processes (Abbott and Nelson 2000; Turrigiano et al. 1998) and relaxation would average synaptic strengths, erasing LTM, LTP, or LTF. We focus here on proposed mechanisms for maintenance of NMDA-dependent, excitatory LTP. Mechanisms for long-term plasticity at other synapse types may differ (Freeman 2015; Hawkins and Byrne 2015; Hige 2018; Lamsa and Lau 2019).

Essential kinases for LTP induction include protein kinase A (PKA) (Abel et al. 1997; Matsushita et al. 2001), the atypical PKC denoted protein kinase M ζ (PKMζ) (Sacktor et al. 1993), the MAP kinase (MAPK) isoform ERK (Rosenblum et al. 2002), and CaM kinase II (CaMKII) (Malenka et al. 1989; Otmakhov et al. 1997). For CaMKII and PKMζ, activation maintained by positive feedback has been hypothesized to be necessary for maintaining L-LTP and LTM. Smolen et al. (2019) discuss hypothesized positive feedback involving ERK and other kinases. Here we limit our focus to CaMKII and PKMζ.

CaMKII. This feedback loop relies on self-sustaining autophosphorylation of CaMKII holoenzymes resulting in persistent autonomous activation (Lisman and Goldring 1988), which has generated, in models, a bistable switch of CaMKII activity and synaptic strength (Aslam et al. 2009; Miller et al. 2005; Zhabotinsky 2000). Active CaMKII can enhance translation (Atkins et al. 2004; Atkins et al. 2005). Persistence of CaMKII activation, and the necessity of CaMKII activation for maintaining L-LTP and LTM, has been challenged (Lengyel et al. 2004; see Smolen et al. 2019 for review). For example, a light-inducible inhibitor of CaMKII was used by Murakoshi et al. (2017) who argued persistent CaMKII activity is not necessary to maintain LTP. However, this study did not specifically examine whether CaMKII inhibition blocks persistence of L-LTP at late times (> 2 h post-stimulus) or late LTM (~24 h post-training). Thus it is still possible that autonomously active CaMKII, perhaps in a small postsynaptic zone, is required to maintain L-LTP and LTM. Saneyoshi et al. (2019) delineated a positive feedback loop that could contribute to localized, persistent CaMKII activation. CaMKII phosphorylates Tiam1, a Rac guanine nucleotide exchange factor. Binding of Tiam1 to CaMKII is thereby enhanced, and Tiam1, in turn, occludes the autoinhibitory domain of CaMKII in a manner that enhances autonomous activity of CaMKII. A caveat is that the activity of CaMKII was not monitored for more than 40 min post-stimulus, preventing direct application of this study to maintenance of L-LTP. A study using dominant negative (dn) CaMKII also suggests persistent CaMKII activity is required to maintain spatial LTM (Rossetti et al. 2017), although with a caveat that the molecular mechanism by which dnCaMKII disrupts LTM is not known and may not depend on activity *per se*. Considering all these data, it remains plausible that persistently enhanced CamKII activity is required to maintain some forms of L-LTP and LTM, although further study is needed.

PKMζ. This positive feedback loop relies on a self-sustaining increase in translation, and thus activity, of PKMζ. Inhibiting PKMζ blocks maintenance of L-LTP and several forms of LTM (Pastalkova et al. 2006; Serrano et al. 2005, 2008; Shema et al. 2007). Following spatial memory formation, PKMζ activity is increased for at least a month (Hsieh et al. 2017). PKMζ upregulates its own synthesis (Kelly et al. 2007a), forming a positive feedback loop. Modeling has suggested this loop can generate a bistable switch (Helfer and Shultz 2018; Jalil et al. 2015; Smolen et al. 2012). As output, PKMζ promotes AMPAR trafficking to synapses (Migues et al. 2010), which could maintain increased synaptic strength. PKMζ also phosphorylates the histone acetyltransferase denoted CREB binding protein, which may enhance transcription required for late maintenance of LTP and LTM (Ko et al. 2016). However, the requirement for persistently enhanced PKMζ activity has been questioned by observations of normal LTP and LTM following constitutive (Lee et al. 2013) or conditional (Volk et al. 2013) PKMζ knockouts. But another study (Wang et al. 2016) has, in contrast, found impaired maintenance of LTM and LTP following knockdown of PKMζ. Thus, it is plausible that persistently enhanced PKMζ activity is necessary for the maintenance of some forms of L-LTP and LTM. But if PKMζ activity is persistently enhanced, is it due to autonomous positive feedback, or is it maintained due to another ongoing process such as synaptic reactivation or persistent CamKII activity?

<u>Reactivation.</u> Specific assemblies of neurons, or memory engrams, are activated both upon learning and during memory retrieval (Park et al. 2016; Zhou et al. 2009) and specific patterns of functional connectivity among neurons within an engram, based on potentiated synapses as well as altered neuronal excitability, are important for memory storage (Choi et al. 2018; Josselyn et al. 2017; Josselyn and Tonegawa 2020; Mozzachiodi and Byrne 2010; Ryan et al. 2015; Tonegawa et al. 2018). Recurrent reactivation of neurons within engrams encoding recent experience, commonly termed replay, appears generic (Ikegaya et al. 2004; Miller et al. 2014; Wu and Foster 2014) and correlates with improved LTM (Schapiro et al. 2018).

Such reactivation of engram synapses may be necessary to maintain potentiated synapses and LTM. For example, inducible forebrain-restricted knockdown of the NR1 subunit of the NMDA receptor, several months post-training, disrupts retention of remote cued and contextual fear memories and of taste memory (Cui et al. 2004, 2005; see also Shimizu et al. 2000). An implication of NR1 necessity is that recurrent reactivation of engram neurons and synapses connecting them induces $Ca^{2+}$ influx through NMDA receptors and voltage-dependent $Ca^{2+}$ channels, thereby reactivating processes, such as kinase activation, that originally induced LTP. By these processes, reactivation could generate recurring "rounds" of LTP, reinforcing and preserving L-LTP and LTM.

In this scenario kinases would, to some degree, be persistently or recurrently active, and make essential contributions in maintaining LTM. However, in contrast to the feedback loops discussed previously, persistent activation of CaMKII, PKMζ, or other kinases would not be self-sustaining. Synaptic reactivation, and consequent kinase activation, would also likely induce transcription *via* stimulus-responsive factors such as cAMP response element binding protein (CREB) (Smolen et al. 2019). Resultant global increases in synaptic proteins may be a required permissive factor for reinforcement of L-LTP.

Summing up the above, evidence suggests self-sustaining kinase activation, in particular persistently elevated activities of CaMKII or PKMζ, may be <u>necessary</u> for the maintenance of LTM. There is also evidence that synaptic reactivation may be necessary. But a further question is, are any of these positive feedback mechanisms also <u>sufficient</u> for maintenance? A problem in addressing the issue of sufficiency is that these positive feedback loops are interlocked such that manipulation of one process affects others. One way to address this issue is to develop a mathematical model of the processes and modify the parameters of each process, and thereby examine conditions under which each process is necessary and sufficient for LTM.

We developed four variants of a simplified model, to compare qualitative dynamics of kinase activities and synaptic strength predicted by these positive feedback loops. Two interrelated but contrasting hypotheses were continued: 1) autonomous activation of one or more kinases such as PKMζ or CaMKII is necessary and sufficient to maintain L-LTP and LTM, and 2) persistent kinase activation is necessary, but it depends on ongoing synaptic reactivation, which is sufficient. Simulations of these mechanisms suggest experiments that can examine dynamics predicted by these hypotheses.

In two model variants, bistable switches of kinase activation are generated respectively by 1) persistently increased PKMζ synthesis and activity, or 2) persistent CaMKII activation. In a third variant bistability in synaptic weight is generated by positive feedback in which synaptic reactivation preferentially occurs at strengthened synapses. L-LTP maintenance requires recurrent kinase activation downstream of reactivation. A fourth variant illustrates that synaptic reactivation and autonomous kinase activation could synergistically maintain L-LTP.

## Model Development

The model describes the dynamics of kinases necessary for L-LTP induction, and represents kinase-dependent setting of a synaptic "tag" that has been shown to be necessary for localized "capture" of plasticity-related proteins necessary for L-LTP (Frey and Morris 1997, 1998). This model is simplified from those we have previously published (Smolen et al. 2006, 2014) in that it focuses on synapse-specific positive feedback loops. It does not represent gene induction and transcriptional or epigenetic positive feedback, although these have been hypothesized as permissive factors for maintaining memory (Smolen et al. 2019; Zhang et al. 2016).

Figure 1 schematizes the model. Stimuli, such as a tetanus, are represented by concurrent brief elevations of synaptic $Ca^{2+}$ and cAMP, and brief activation of Raf kinase upstream of ERK. Stimulus parameters, including durations, are provided in Methods and are similar to Smolen et al. (2014). Some intermediate processes, such as interaction of $Ca^{2+}$ with calmodulin or calmodulin with CaMKII, and Ras activation upstream of Raf, are not included in the model. CaMKII, PKA, and ERK then activate a synaptic tag to place a synapse in a labile state capable of "capturing" plasticity factors (proteins or mRNAs) and incorporating them to increase synaptic strength. Parameters describing kinetics of enzyme activation are given in Methods. Data support the importance of these kinases for tag setting. At least one tag modification appears to be mediated by PKA (Barco et al. 2002). Postsynaptic CAMKII activity is required for L-LTP, and tag setting is blocked by inhibition of CaMKII (Redondo et al. 2010). CaMKII stimulates translation via phosphorylation of cytoplasmic polyadenylation element binding protein, a candidate for Tag-1 (Atkins et al. 2004, 2005). ERK is also necessary for LTP (English and Sweatt 1997; Rosenblum et al. 2002). In the model, tagging requires phosphorylation of three sites; Tag-1, Tag-2, and Tag-3; respectively substrates of CAMKII, PKA, and synaptic ERK. We assume a multiplicative dependence of synaptic tagging on these phosphorylations, thus the tag (variable TAG) is the product of amounts of phosphorylated Tag-1 – Tag-3. The molecular nature of the synaptic tag is not characterized, therefore this multiplicative dependence is an assumption. However, this dependence is consistent with the block of tag setting by inhibition of a single kinase, CaMKII (Redondo et al. 2010), and with the requirement of ERK activity for LTP (Rosenblum et al. 2002).

CaMKII and ERK are assumed to cooperate to induce PKMζ synthesis, because inhibition of these kinases impairs PKMζ translation (Kelly et al. 2007). PKMζ translation is increased by LTP induction (Hernandez et al. 2003), and PKMζ concentrates at potentiated synapses (Hsieh et al. 2017). Stimulus-induced increases in TAG and PKMζ are assumed to converge, with these variables multiplying to determine the rate of increase of synaptic weight, denoted W. Increased W corresponds to L-LTP. Without positive feedback, W decays to baseline with a time constant of several hours. We include as a parameter the level of a generic plasticity-related protein (PRP) required for LTP. A multiplicative dependence for the rate of increase of W is assumed, with this rate proportional to PRP, TAG, and PKMζ. This multiplicative dependence is a model assumption and is a phenomenological representation of the net effect of numerous processes. It is chosen to represent the separate necessity of each of the components (TAG, PRP, and PKMζ) which must act in concert to support L-LTP induction. There is also a small basal rate of increase in W, to yield a positive basal W in the absence of stimulus.

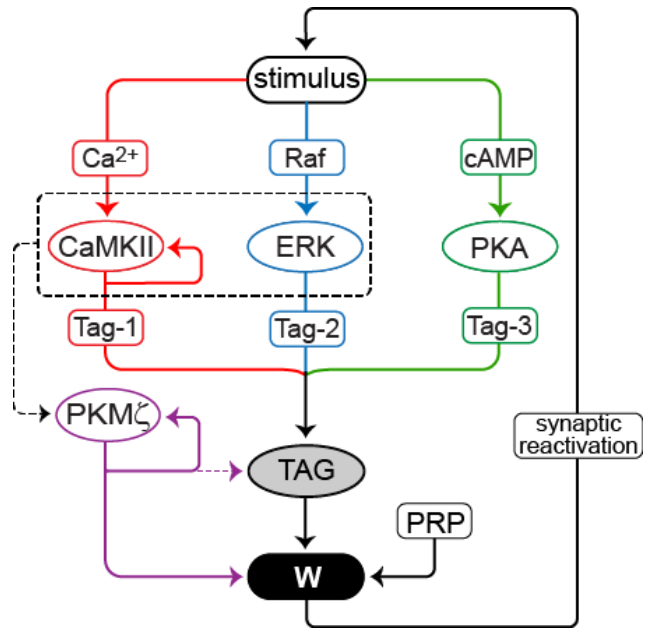

FIGURE 1. **Model for maintenance of LTP. Three model variants contrast positive feedback loops mediated by CaMKII, PKMζ, and synaptic reactivation.** Stimuli activate CaMKII, PKA, Raf, and ERK. CaMKII, ERK, and PKA phosphorylate a synaptic tag. CaMKII, and ERK also converge (dashed box and arrow) to induce translation of PKMζ. Production of plasticity-related protein (PRP) is necessary to maintain LTP. TAG, PRP, and PKMζ converge to increase synaptic weight W. The feedback loops are: 1) PKMζ can enhance its own translation and thus activity. For this model variant, two cases are simulated: a) PKMζ contributes to increasing W, but not to maintaining TAG, and b) PKMζ also contributes to maintaining an elevated level of TAG (dashed arrow). 2) Active CaMKII can reinforce and maintain its own activity. 3) Ongoing synaptic reactivation. An increase in W enhances the amplitude of spontaneous LTP-reinforcing stimuli. Only for potentiated W is the stimulus amplitude sufficient to significantly reactivate downstream signaling pathways.

Three forms of positive feedback are represented in distinct model variants, I – III respectively. Each type of feedback is able, on its own, to generate robust bistability in W, maintaining L-LTP. Simulations consider one type of positive feedback at a time. These feedback loops (Fig. 1) are: 1) Self-sustaining synthesis of PKMζ. Persistent activity is implemented as a Hill function in which active PKMζ feeds back to increase the rate of its own synthesis, with this activation necessary and sufficient for maintaining increased W. This simplified feedback could serve as a proxy for more detailed mechanisms, such as repression by PKMζ of the translation inhibitor Pin1 (Westmark et al. 2010). 2) Self-sustaining CaMKII autoactivation. Persistent activity is implemented as a Hill function in which active CaMKII feeds back to increase the rate of its own activation. This simplified feedback could serve as a proxy for mechanisms including incorporation and activation of new subunits in a CaMKII holoenzyme (Miller et al. 2005) or reciprocal activation of CaMKII and Tiam1 (Saneyoshi et al. 2019). 3) Spontaneous reactivation in which a potentiated synapse is subject to brief, repeated elevations of $Ca^{2+}$ and cAMP. Long-lasting kinase activation requires synaptic reactivation. Reactivation amplitudes are assumed to be greater at stronger synapses (i.e., the amplitudes of these brief $Ca^{2+}$ and cAMP elevations increase with synaptic weight W). Therefore reactivation is only effective at increasing W, counteracting its spontaneous decay, at synapses that are already strong. It is also assumed that reactivation stimuli, which only need to generate minor increases in W to counteract slow passive decay, are weaker than the stimulus that induces L-LTP. Equations and parameters are given in Methods.

To maintain L-LTP, TAG must remain elevated to some extent, either periodically (due to synaptic reactivation) or constitutively. TAG elevation allows a slow ongoing incorporation of plasticity-related protein, increasing synaptic weight to counter passive weight decay. Empirically, the tag is observed to decay within ~ 1-2 h after LTP induction (Frey and Morris, 1997, 1998). But in our simulations, the elevation of TAG required during LTP maintenance remains well below the peak value of TAG during LTP induction. Therefore the model may not conflict with these data (see also Discussion).

## Results

**Simulated L-LTP maintenance due to self-sustaining kinase activation.**

L-LTP induction. We first turned off all positive feedback loops, simulating only induction of L-LTP (Fig. 2A1-A2). We simulated a standard induction protocol – three 1-sec tetani with interstimulus intervals of 5 min. CaMKII responds rapidly to each $Ca^{2+}$ increase and deactivates in ~1 min, similar to the time course in dendritic spines (Lee et al. 2009), whereas synaptic ERK activity, and PKA activity, accumulate over tetani. The synaptic tag variable TAG increases by a large relative factor from basal to peak, due to the multiplicative convergence of ERK, PKA, and CaMKII activities to increase TAG. In alternate simulations in which an additive kinase interaction was assumed, the relative increase in TAG was much lower, resulting in excessive basal levels and in a lack of effective block of LTP upon single kinase inhibition (not shown).

The synaptic weight W increases for > 1 h after stimulus, consistent with data illustrating that L-LTP takes > 1 h to develop when induced by chemical stimuli that bypass early LTP (Ying et al. 2002). The magnitude of L-LTP was assessed 2 h after the last tetanus, as the increase in W relative to pre-stimulus basal level. Standard values for all parameters yielded L-LTP of 131%. Empirical excitatory postsynaptic potential increases after three or four tetani have similar values (English and Sweatt 1997, Woo et al. 2000). Without positive feedback, L-LTP decayed with W returning to near basal ~ 10 h post-stimulus. This time scale for decay of W is set by the time constant $\tau_{ltp}$ (Methods, Eq. 21), and was chosen to allow comparison of model dynamics when L-LTP was maintained by the feedback mechanisms proposed above.

Self-sustained elevation of PKMζ synthesis. In Fig. 2B1-B2, this positive feedback loop sustains a bistable switch, with PKMζ persistently active. All other dynamic variables, including TAG and W, decay to basal levels post-stimulus. L-LTP is not maintained. This lack of maintenance follows from the lack of dependence of TAG on PKMζ phosphorylation. Without elevated TAG, plasticity-related protein (PRP) cannot be incorporated. Thus an important implication, and topic for future empirical investigation, is that if self-sustained activity of PKMζ is sufficient to maintain LTP, then either: 1) the tag is not needed for maintenance, or 2) PKMζ activation participates in and is sufficient for persistent synaptic tagging, or 3) there is a constitutive basal level of the synaptic tag, independent of kinase activation during L-LTP induction and sufficient to maintain L-LTP. In cases 2) and 3), as noted above, the low level of TAG required for maintenance may not conflict with empirical observations of synaptic tag decay. Activation of metabotropic glutamate receptors could allow PKMζ to participate in synaptic tagging (Li et al. 2014)

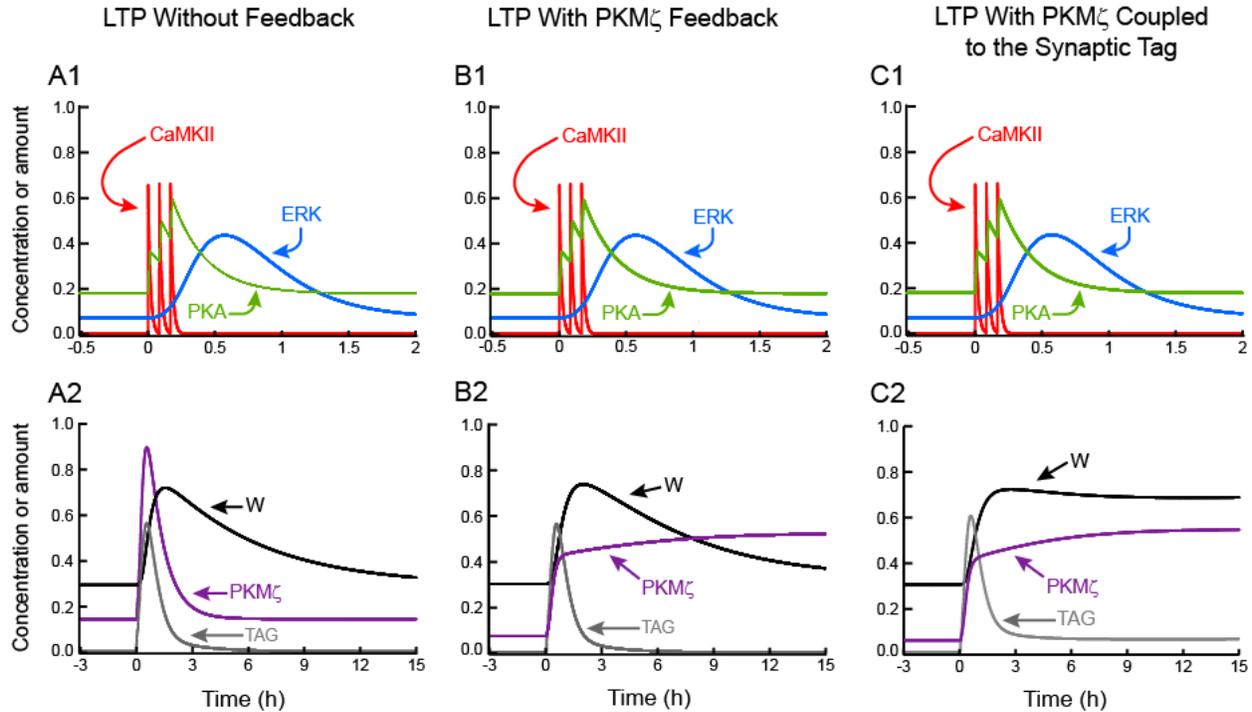

FIGURE 2. **Simulated LTP, without and with positive feedback based on sustained PKMζ activity.** **(A1)** A three-tetanus stimulus induces peaks of CaMKII activity and elevates activities of PKA and ERK. Without positive feedback, kinase activities return to basal with the slowest variable, ERK, elevated for ~1.5 h. **(A2)** Increased kinase activities activate TAG and PKMζ. LTP decays over h. **(B1)** With persistent PKMζ activation, a tetanic stimulus leads to only temporary activation of CaMKII, ERK, and PKA. **(B2)** Enhancement by PKMζ of its own synthesis and activity leads to bistability and a state switch for PKMζ. But TAG is not persistently elevated and LTP is not maintained. **(C1) – (C2)** If PKMζ is assumed to increase the level of TAG, then stimulus-induced persistent PKMζ activation leads to elevation of TAG and maintenance of LTP. In Figs. 2 - 5, time courses of some variables are vertically scaled for ease of visualization. Scaling factors are given in Methods.

To illustrate maintenance of L-LTP dependent on activation of the synaptic tag by PKMζ, we repeated the simulation of Fig. 2B with the following modification. The equation giving the level of TAG (Eq. 17, Methods) was modified by adding a positive term to the right-hand side, $k_{TPKM} PKM$. This is the simplest plausible dependence. It assumes the level of TAG and the activity of PKMζ equilibrate rapidly compared to changes in PKMζ. As illustrated in Fig. 2C1-C2, this PKMζ dependence suffices for bistability, with persistent maintenance of L-LTP. During maintenance, minor but persistent elevation of TAG above its pre-stimulus baseline is evident. This elevation is minor because the other kinases that contribute to tag setting (CaMKII, ERK, PKA) are not persistently activated. Minor TAG elevation suffices to allow continued slow incorporation of plasticity-related protein to counter passive synaptic weight decay. This minor TAG elevation is much less than the peak of TAG during L-LTP induction. It may not be readily evident empirically, and thus may not conflict with data suggesting a tag lifetime of 2-3 h (Frey and Morris 1997, 1998). In this simulation, for bistability in W and TAG to be evident, it is essential to assume that PKMζ activity increases TAG additively, with the $k_{TPKM} PKM$ term independent of CaMKII, ERK, and PKA. If, instead, PKMζ activity simply multiplies together with the activities of CaMKII, ERK, and PKA, then the persistent elevation of TAG and W is negligible, because of the return of CaMKII activity to its very low basal value.

The model can also simulate maintenance of L-LTP for the case in which a low, basal level of synaptic tagging is assumed to persist, independent of the activity of all modeled kinases (not shown). In this case, the maintained increase in synaptic weight W is entirely due to the persistent increase in PKMζ activity.

<u>Self-sustained autoactivation of CaMKII.</u> To simulate maintenance of L-LTP due to self-sustained activation of CaMKII, PKMζ autoactivation was removed and feedback mechanism 2 was implemented (Methods, Eq. 23). In Fig. 3A1-A2, this feedback sustains a bistable switch, with CaMKII activity and W persistently active. ERK and PKA activities still decay. Similarly to Fig. 2C2, TAG remains persistently elevated. This elevation is again well below the peak during induction. Intriguingly, PKMζ activity also remains elevated. CaMKII increases synthesis and activity of PKMζ (Kelly et al. 2007). Thus in this model variant, PKMζ activity depends on a CaMKII-catalyzed phosphorylation (Eqs. 18 and 20, Methods), so that both activities are concurrently enhanced. PKMζ activity, although persistent, is not autonomously self-sustaining. However, as noted in the Introduction, the hypothesis that autonomous CaMKII activity sustains L-LTP has substantial empirical challenges.

**Simulated synaptic reactivation maintains L-LTP by reactivating kinases and tagging.**

Maintaining L-LTP by ongoing reactivation of a strengthened synapse (feedback mechanism 3) yields qualitatively different dynamics. As illustrated in Fig. 3B1-3B2, recurrent increases of cAMP, $Ca^{2+}$, and Raf activity are induced by reactivations (Methods, Eqs. 25-27). These increases activate ERK, PKA, and CaMKII, synthesize PKMζ, and recurrently elevate the synaptic tag. These increases were adjusted by trial and error to maintain a stable, although fluctuating, state of increased synaptic strength. During reactivation, recurrent peak elevations of kinase activities excepting CaMKII, and of the synaptic tag, are substantially less than during LTP induction, and TAG relaxes to low levels between reactivations. If, as assumed here, reactivations are not very frequent, then these dynamics may be compatible with empirically reported decay of the tag within ~1-2 h after LTP induction. These simulations illustrate a model property more general than this specific reactivation loop. Sustained or recurrent activation of all simulated kinases is necessary and sufficient to maintain a degree of synaptic tagging and maintain L-LTP.

The simulation of Fig. 3B1-B2 thus indicates a way in which LTP could be maintained in the absence of autonomous biochemical feedback loops. One qualitative difference between this maintenance mechanism and those based on autonomous kinase activation is that here, the synaptic weight can exhibit significant oscillations depending on the frequency of reactivation, whereas maintained weights may be more stable with automous kinase activation.

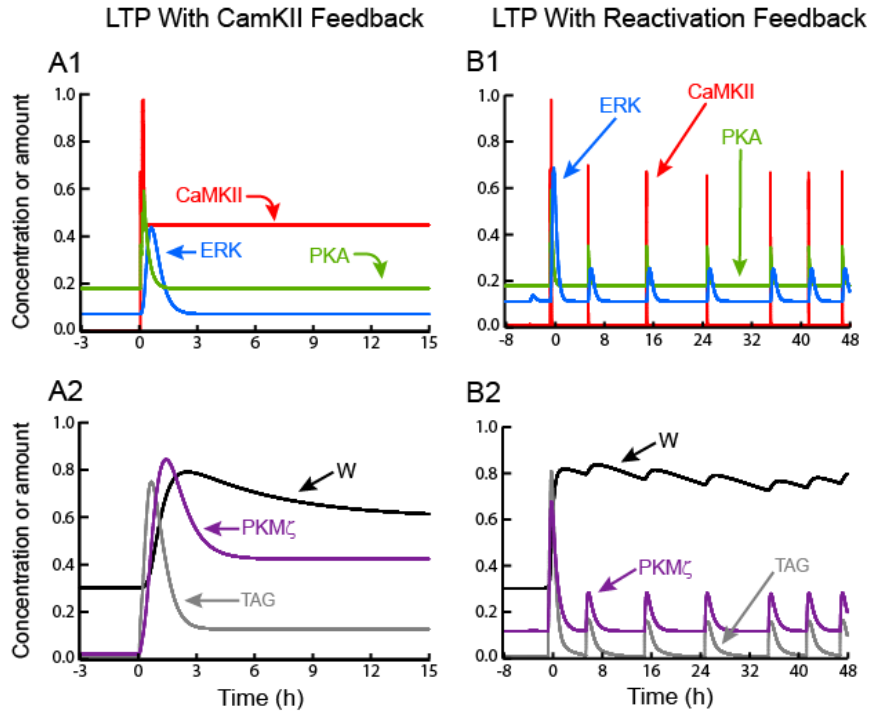

FIGURE 3. **Simulated LTP with positive feedback based on sustained CaMKII activity or on synaptic reactivation.** **(A1)** and **(A2)** LTP maintenance due to autoactivation of CaMKII. **(B1)** and **(B2)** Maintenance due to positive feedback in which a strengthened synapse undergoes reactivation events, with each event reinforcing LTP. In both these model variants, PKMζ does not act to increase the level of TAG.

**The model is not unduly sensitive to most parameter variations.**

To determine if the model variants discussed above are overly sensitive to parameter changes, the sensitivity of the stimulus response – the amount of L-LTP – to relatively minor changes (±15%) of all parameter values including stimulus amplitudes was analyzed, perturbing from the standard values (Methods). There are 43 parameters, yielding 86 parameter variations for each model variant (note for each variant, 2-3 parameters governing strength of positive feedback are set to zero, thus these parameter variations are zero). We used a standard analysis defining a set of relative sensitivities $S_i$, with i ranging over all parameters $p_i$ (Frank, 1978). R denotes the magnitude of stimulus response. For each $p_i$, a change is made, and the resulting change in R is determined. Relative sensitivity $S_i$ is defined as the absolute value of the relative, or fractional, change in R divided by the relative change in $p_i$,

$$S_i = \left| \frac{\Delta R / R}{\Delta p_i / p_i} \right| \qquad 1)$$

R is the magnitude of L-LTP after tetani. The sensitivity of four simulations was assessed – without feedback (Fig. 2A), with kinase feedback (Figs. 2B and 3A), and with reactivation feedback (Fig. 3B). For Figs. 2A and 2B, which were not bistable in W, L-LTP was quantified 2 h post-stimulus, whereas for the bistable simulations of Figs. 3A and 3B, L-LTP was quantified 50 h post-stimulus.

Sensitivity to most of the 86 parameter variations was not large, with $S_i$ below 3. There were 13 exceptions with $S_i$ between 3.0 and 9.9. Interestingly, 10 of these exceptions were for Raf-ERK pathway parameters (Methods, Eqs. 6-13), and these 10 were >3 for all four simulations. These exceptions were: 15% increases and decreases in the total concentration of Raf kinase ([$Raf_{tot}$], Eq. 6 in Methods), in the rate constant for Raf activation ($k_{fbasRaf}$, first paragraph of Methods and Eq. 7), in the rate constants for MEK activation ($k_{fMEK}$) and inactivation ($k_{bMEK}$), and in the inactivation rate constant for Raf ($k_{bRaf}$). Two other exceptions were for ERK rate constants (+$k_{fERK}$, -$k_{bERK}$). Finally, the simulation with PKMζ autoactivation (Fig. 2B) had $S_i = 7.0$ for a decrease in the corresponding Hill constant ($K_{pkm}$, Eq. 24). This lack of unduly large sensitivity for most parameter variations (73 out of 86) suggests that model variants I - III are relatively robust and can be regarded as reasonable qualitative models of synaptic positive feedback mechanisms. Further studies might seek to add buffering mechanisms, especially within the Raf-ERK pathway, to reduce sensitivity to the specific parameters noted.

**Simulations suggest a test to distinguish feedback mechanisms for maintenance of L-LTP.**

Further simulations suggest the consequences of inhibition of key kinases, in particular PKMζ or CaMKII, may provide an empirical test to distinguish between feedback mechanisms. Fig. 4A illustrates that if self-sustained activation of CaMKII is sufficient to maintain L-LTP, then transient inhibition of a different kinase, in this case PKMζ, would not abolish L-LTP maintenance. W repotentiates after inhibition is removed. Repotentiation occurs because CaMKII autoactivation does not require PKMζ activity and persists throughout inhibition. In contrast to this simulation, empirical inhibition of PKMζ does appear to inhibit L-LTP and LTM (Ling et al. 2002; Serrano et al. 2008; Shema et al. 2007), suggesting self-sustained CaMKII activation is not sufficient. However, CaMKII autoactivation may be necessary, as some data suggests (Rossetti et al. 2017; Sanhueza et al. 2007) and as we have simulated (model variant II).

Figure 4B illustrates that if recurrent synaptic reactivation maintains L-LTP, then inhibition of PKMζ can permanently reverse L-LTP. W declines while PKMζ is inhibited, because PKMζ activity is needed to maintain elevated synaptic weight. Therefore, if W declines sufficiently during PKMζ inhibition, selective reactivation of a strong synapse is no longer present, and LTP cannot be maintained. Consequently, even if inhibition of PKMζ is then removed, W continues to decline towards its basal value (Fig. 4B). For standard parameter values (Methods), PKMζ inhibition does need to last ~ 50 h in order for W to decrease sufficiently, so that after inhibition removal reactivation is not able to re-potentiate W. This simulation argues that empirically, inhibition of PKMζ during the maintenance phase of L-LTP does not enable discrimination between: A) the sufficiency of a feedback loop based on biochemically driven, activity-independent PKMζ activation, *vs.* B) the necessity, or even sufficiency, of a reactivation feedback loop involving PKMζ as an essential component.

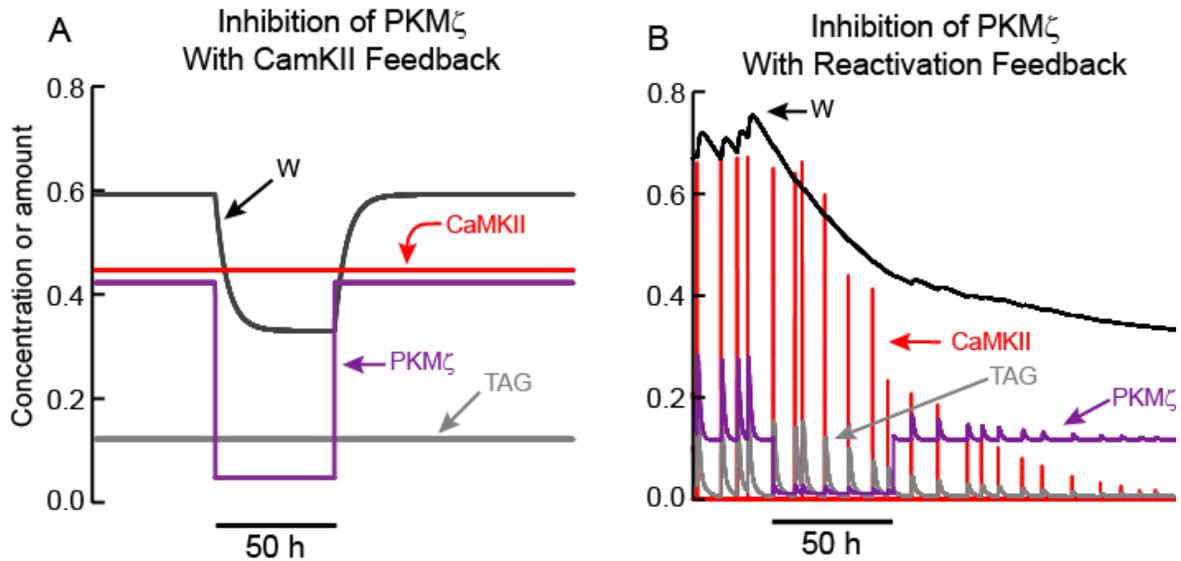

FIGURE 4. **Simulated effects of PKMζ inhibition on LTP maintenance.** (**A**) When LTP is maintained by CaMKII activation, only a temporary reduction in synaptic weight occurs. TAG does not depend on PKMζ and remains elevated. Thus, following PKMζ inhibition, TAG and PKMζ can again cooperate to increase W, restoring LTP. (**B**) With positive feedback based on reactivation of strong synapses, a decrease in W, induced by PKMζ inhibition, decreases amplitudes of reactivation stimuli and activation of CaMKII and other kinases, decreasing TAG. Long-lasting inhibition of PKMζ (90% inhibition for 50 h) decreases W below a threshold, such that the bistable system transitions to the basal state with low W, PKMζ, and TAG. In this model variant, one slow time constant is assumed to describe synaptic weight decay. Because of this slow time constant, the inhibition duration required to decrease W below threshold is ~ 2 d. Empirically, PKMζ inhibition may reduce synaptic weight more rapidly than would a simple loss of reactivation, in which case a shorter inhibition duration would suffice.

**Synaptic reactivation and autonomous kinase activation could act synergistically to maintain L-LTP.**

The preceding model variants contrast qualitative dynamics that characterize maintenance of L-LTP and LTM due to one of two plausible forms of positive feedback, autonomous kinase activation and synaptic reactivation. It is also possible that multiple positive feedback loops act in concert to sustain bistability in synaptic weight, maintaining L-LTP and LTM. To explore dynamics associated with this scenario, we constructed a fourth model variant (Methods). We modified model variant III, based on synaptic reactivation, to include self-sustaining enhancement of PKMζ translation as in variant I. Parameters determining the strength of both forms of feedback were reduced from their standard values, so that neither reactivation nor PKMζ activation alone could maintain L-LTP. However, when both forms of positive feedback operated concurrently, bistability resulted in PKMζ activity and in W, and L-LTP was maintained.

Figure 5A1-A2 illustrates induction of L-LTP and its maintenance post-tetanus. These dynamics are similar to Fig. 3B. During L-LTP maintenance, activities of CaMKII, ERK, and PKA, along with the synaptic tag, fluctuate above basal levels. At 2000 min post-stimulus PKMζ positive feedback was switched off. W, and PKMζ activity, both decline to basal levels (with a slower time constant for W), illustrating that reactivation alone cannot sustain bistability.

Figure 5B illustrates an analogous simulation, for which the PKMζ positive feedback remains on and reactivation is switched off. Bistability is again lost, and PKMζ activity and W both decline to basal levels, although the decline is quite slow for the simulated parameter values.

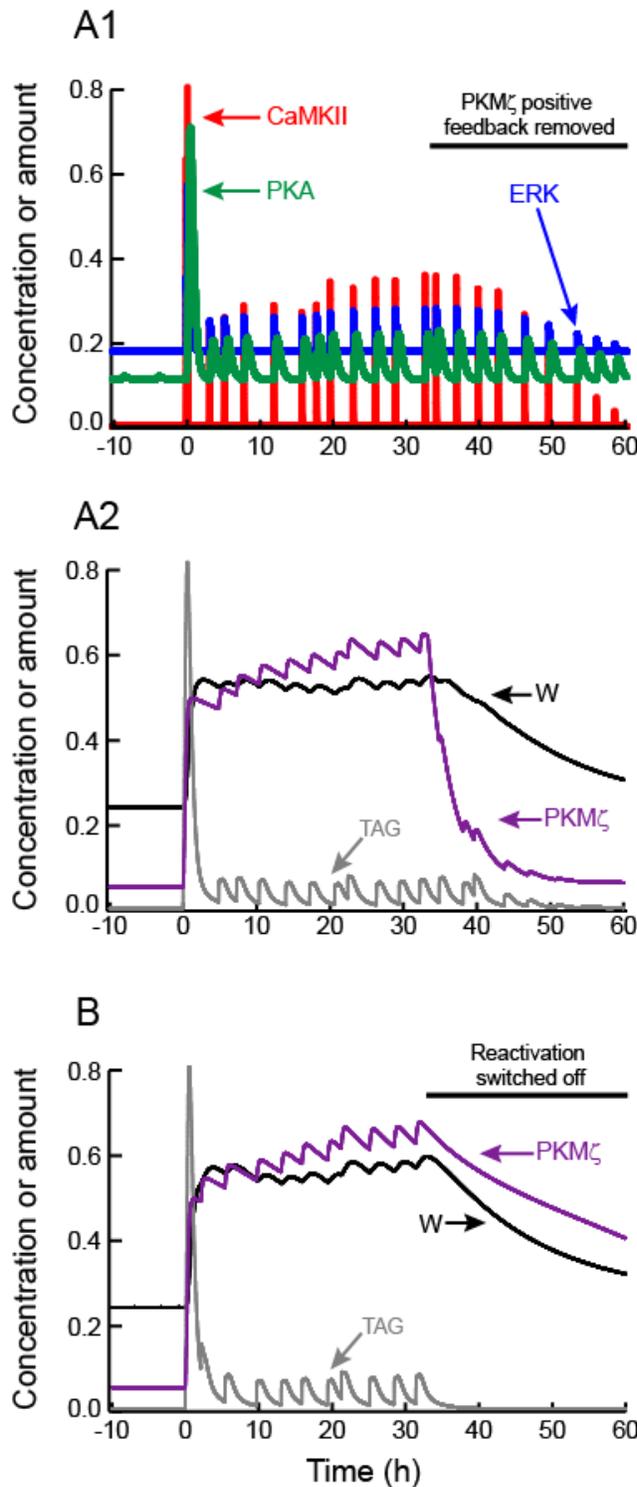

FIGURE 5. **Maintenance of L-LTP by mutual reinforcement of self-sustaining PKMζ activity and synaptic reactivation.** **(A1)** and **(A2)** Tetanic stimulus induces kinase activation, synaptic tagging, and persistent increase in W. After W increases, reactivation leads to recurrent increases in CaMKII, ERK, and PKA activities in **A1**, and TAG in **A2**, but these variables remain well below peak values during LTP induction. Starting at $t = 33$ h (black horizontal bar), PKMζ positive feedback is switched off by setting the parameter $k_{PKM}$ to zero. Despite continued synaptic reactivation, PKMζ and W decline to basal levels. **(B)** LTP is induced and maintained with the same parameters as in **A**, with only the random timing of reactivation events differing. Starting at $t = 33$ h (black horizontal bar), reactivation is switched off, while PKMζ positive feedback remains on. PKMζ and W decline slowly to basal levels

## Discussion

We have considered two interrelated hypotheses: 1) autonomous self-sustaining activation of a kinase, focusing on PKMζ or CaMKII, is necessary and sufficient to maintain L-LTP; and 2) persistent activation of kinases is not autonomous, and is recurrently induced by synaptic reactivation, which maintains L-LTP. There is empirical data in favor of each hypothesis. Modeling investigated whether the dynamics of kinase activities, synaptic tagging, and synaptic weight predicted by these hypotheses could be discriminated. We identified dynamic differences, and make empirical predictions that could further discriminate between these mechanisms.

### Dynamics of autonomous kinase activity

L-LTP was maintained by the positive feedback loop based on autonomous CaMKII activation (Fig. 2C), or the loop based on synaptic reactivation (Fig. 3A), or the loop based on autonomously increased synthesis of active PKMζ (Fig. 2C). With autonomous CaMKII activation, CaMKII activity remains high during maintenance, well above the other kinases. In contrast, when L-LTP is maintained by persistent PKMζ activation or by synaptic reactivation, CaMKII activity is maintained at a much lower level, sufficient to sustain, in cooperation with the other kinases, a low persistent level of TAG that allows continued necessary incorporation of plasticity-related proteins (PRPs). In these latter two cases, low CaMKII activity suffices because its cooperating kinase, PKMζ, is persistently or recurrently elevated.

Model variant I, based on autonomous, self-sustaining synthesis of active PKMζ did not initially maintain L-LTP (Fig. 2B), because we first assumed there was no persistence of the synaptic tag considered essential for ongoing incorporation of plasticity related proteins (PRPs). However, when PKMζ activity was assumed to elevate TAG, L-LTP was maintained (Fig. 2C). With this model variant, L-LTP could also be maintained if a low basal level of TAG, independent of the kinase activities in the model, was assumed to persist.

If a positive feedback loop based on intrinsic activation of a kinase, such as CaMKII or PKMζ, is necessary and sufficient for maintenance of L-LTP, then the simulation of Fig. 4A illustrates that inhibition of a kinase or process not within the feedback loop should fail to permanently reverse L-LTP. If sustained activity of PKMζ is both necessary and sufficient, then inhibiting CaMKII should have no effect, and *vice versa*. Assuming inhibitor specificity, the empirical loss of established L-LTP and LTM upon PKMζ inhibition implies no feedback loop (intrinsic or network based) independent of PKMζ can be both necessary and sufficient for maintenance.

We note additional positive feedback loops have been proposed to maintain LTM (Smolen et al. 2019). For example, in mammals and in *Aplysia*, isoforms of the translation activator denoted cytoplasmic polyadenylation element binding protein (CPEB) can upregulate translation at synapses that have undergone LTP or LTF. Aggregation of CPEB and increased translation has been hypothesized to be self-perpetuating and to maintain synaptic strength (Drisaldi et al. 2015; Fioriti et al. 2015; Li et al. 2018; Rayman and Kandel 2017; Si et al. 2003; 2010). Our model variants have not considered self-sustaining CPEB aggregation. However, data suggesting that ongoing synaptic reactivation is necessary in mammals (Cui et al. 2004, 2005; Shimizu et al. 2000) imply that autonomous CPEB aggregation would not be sufficient.

We have not further explored with simulations the possible consequences of a low basal TAG independent of kinase activities. However, it is qualitatively evident that in this case, any persistent activation of a self-sustaining kinase or other process that increases PRP levels could maintain L-LTP, *via* enhanced PRP incorporation driven by basal TAG. For example, if CPEB aggregation increased translation of PRPs, then a basal TAG level could suffice to facilitate incorporation of these PRPs, maintaining L-LTP. Persistent CaMKII activity could upregulate PRP translation via CPEB phosphorylation (Atkins et al. 2004, 2005), maintaining L-LTP.

**Dynamics of synaptic reactivation**

In model variants III-IV for which recurrent synaptic reactivation is essential to maintain L-LTP, only potentiated synapses undergo substantial reactivation to reinforce L-LTP. This assumption needs to be empirically verified. If synaptic reactivation does maintain strong synapses and is stronger or more frequent at potentiated synapses in memory engrams, then a prediction is that enhanced reactivation yields increased time-average activity of kinases, specific to potentiated synapses, and reinforcing LTP. As with LTP induction, these kinases include ERK, CaMKII, PKA, and plausibly PKMζ. Monitoring FRET-engineered substrates of these kinases over relatively long periods (h or more) might demonstrate increased activity. In addition, simulated dynamics of synaptic reactivation (Fig. 3B) differ from those of autonomous kinase activation (Figs. 2C, 3A) in that reactivation events generate large fluctuations in the activities of synaptic CaMKII, ERK, and the level of TAG (Fig. 3B). Such fluctuations may be observable empirically. Our model has not addressed whether the frequency of reactivation events would be increased at potentiated synapses. To model reactivation frequency would require a complex neuronal network model within which a synaptic memory engram can be stored and maintained, and reactivation events can be self-consistently generated.

It would be highly desirable to strengthen the empirical case for the necessity of synaptic reactivation by examining the consequences, for maintenance of L-LTP and LTM, of additional pharmacological or genetic methods for prolonged synaptic activity reduction. Because reported timescales on which reactivation appears necessary are long – weeks to months – such studies will need to be done *in vivo* rather than with *in vitro* preparations of limited duration. In *in vitro* slice preparations, over times of hours, synaptic activity patterns may not recapitulate patterns *in vivo*, reducing the likelihood that a positive feedback loop based on reactivation of specific synapses operates. Biochemical positive feedback, such as persistent kinase activation, may contribute to maintaining L-LTP over hours *in vitro*.

Although the present study has focused on engram-level reactivation, another form of reactivation could be mediated by autocrine actions. For example, brain-derived neurotrophic factor (BDNF) released by synaptic activity can act, via TrkB receptors, to reactivate biochemical pathways necessary for LTP, including the Ras → Raf → MEK → ERK signaling cascade (Panja and Bramham 2014). These events can upregulate transcription of *bdnf*, generating a positive feedback loop (Bambah-Mukku et al. 2014). Our model does not represent dynamics of neurotrophins, but we note that recurrent engram-level reactivation would be expected to upregulate BDNF secretion (Kolarow et al. 2007), with secreted BDNF reinforcing LTP (Barco et al. 2005; Kovalchuk et al. 2002). A modeling study does suggest the positive feedback loop in which BDNF upregulates *bdnf* expression may play a role in maintaining L-LTP (Zhang et al. 2016).

**Predictions to help discriminate between proposed maintenance mechanisms**

The simulation of Fig. 3A2 illustrates that sustained activation can depend on a separate process in a way that complicates determination of which process is necessary. Here, persistent activation of PKMζ occurs, but it is not autonomous. Instead, persistent PKMζ activity depends on self-sustaining activation of CaMKII, which enhances PKMζ synthesis. These dynamics suggest that if CaMKII activation maintains L-LTP, this maintenance could be due in part to downstream PKMζ activation. If so, a prediction is that if CaMKII is inhibited to abolish L-LTP, PKMζ levels at potentiated synapses would simultaneously decay. In contrast, if PKMζ activity is elevated due to an autonomous biochemical feedback loop, then elevated synthesis and activity of synaptic PKMζ may not be reversed by inhibition of CaMKII, or by inhibition of synaptic reactivation.

Simulations of the model variant for which selective reactivation of strong synapses suffices to maintain L-LTP (Fig. 4B) support a prediction that for this mechanism, inhibition of any kinase will permanently reverse L-LTP only if inhibition lasts long enough to decrease synaptic weight such that synapse-specific reactivation no longer occurs, and LTP is not reinforced. In this simulation PKMζ was inhibited, but the same argument applies for inhibition of CaMKII or any other kinase necessary for L-LTP. Prolonged, neuron-specific inhibition of kinases such as ERK, PKA, or CaMKII may provide a test of this prediction.

When we modeled L-LTP maintenance as dependent on synaptic reactivation, elevation of PKMζ activity was still assumed necessary for L-LTP maintenance, but this elevation was due to synaptic reactivation and was not autonomous. For this mechanism, empirical inhibition of synaptic activity would be predicted to knock down PKMζ levels, testing whether synaptic PKMζ elevation is in fact autonomous.

It appears difficult to empirically separate a model in which PKMζ positive feedback and synaptic reactivation are both necessary to sustain L-LTP (Fig. 5) from reliance on reactivation alone (Fig. 3B). The difficulty occurs because for both variants, inhibiting either PKMζ or synaptic reactivation abolishes bistability and maintenance of L-LTP.

Our model is limited in what it can infer concerning the dependence of L-LTP maintenance on protein turnover rates. For CaMKII, our assumed fast deactivation time constant (1 min) is consistent with observed rapid kinetics of CaMKII deactivation (Murakoshi et al. 2017), but is far slower than protein turnover. Half lives of synaptic CaMKII α, δ, and γ isoforms have been measured to be 3 – 4 days (Cohen et al. 2013), and a more complex model would be needed to simulate CaMKII turnover. Such a model would need to represent CaMKII subunit turnover and replacement within a CaMKII holoenzyme, and also account for the likelihood that a persistently active population of postsynaptic CaMKII should be modeled as a separate pool. Singh and Bhalla (2018) have recently modeled turnover and subunit exchange among CaMKII holoenzymes and have simulated bistability, with persistence of enhanced CaMKII activity for at least 60 days. As regards maintenance of L-LTP by recurrent synaptic reactivation (model variant III), this maintenance mechanism is relatively insensitive to protein turnover rates. Kinases, and the synaptic tag, can deactivate or decay relatively quickly, as long as reactivation events occur frequently enough so that reactivation of kinases, and possibly increased translation of postsynaptic proteins, induces periodic elevations of TAG levels and of synaptic weight to counteract ongoing weight decay.

As regards maintenance of L-LTP by bistable PKMζ activity (model variant I) our model assumes a standard value of 0.02 min$^{-1}$ for the PKMζ deactivation rate constant. If deactivation corresponds to degradation, then the degradation time constant is 50 min. Cohen et al. (2013) did not measure the turnover time of synaptic PKMζ, and as far as we know it has not yet been determined. Eden et al. (2011) measured half-lives of ~100 proteins from a human cancer cell line and found values ranging from 45 min to 23 h. These data suggest that if PKMζ deactivation corresponds to degradation, our assumed rate constant is at the upper end of plausible values. With the current model, we are not able to substantially decrease this rate constant. If the PKMζ degradation time is assumed to be substantially longer, e.g., several days as found for the majority of synaptic proteins (Cohen et al. 2013), a more complex model would be needed. Such a model might assume a relatively rapid sequestration of constitutively active PKMζ, or an inactivating modification of PKMζ, followed by a slow degradation.

A basic assumption, common to all our simulations, is that processes necessary for the induction of L-LTP are also necessary for maintenance of L-LTP of engram synapses. This assumption will need to be empirically tested. We assume the synaptic tag necessary for "capture" of plasticity-related proteins (PRPs) during induction of L-LTP remains essential during maintenance of L-LTP to allow reinforcing capture of PRPs. If this assumption is correct, the tag is predicted to be persistently set to some extent, or periodically reactivated, at strong synapses in order to allow reinforcing capture of PRPs. During maintenance, the activities of CaMKII, ERK, and PKMζ are all predicted to be necessary insofar as these kinases participate in tag setting, or enhance synaptic AMPAR incorporation. But autonomous bistability in these activities may (model variants I, II, IV) or may not (variant III) be necessary.

The identity of the tag has yet to be established, but several molecular species have been suggested to participate including the BDNF receptor TrkB (Lu et al. 2011) and polymerized actin (Ramachandran and Frey 2009). One or more of these species is predicted to be persistently, or recurrently, upregulated at strong synapses. But the extent of upregulation during L-LTP maintenance, required to counter passive synaptic weight decay, may be substantially less than the increase during induction of L-LTP (Figs. 2C2, 3A2, 3B2). Thus it may be difficult to empirically demonstrate tag persistence or recurrence. Nonetheless, a promising strategy to further delineate the feedback mechanism(s) responsible for maintaining L-LTP and LTM is to investigate whether, and which, kinase activities or suggested tagging molecules are upregulated selectively at strengthened engram synapses.

## Methods

### Equations prior to positive feedback

In the model, LTP is induced by three tetanic stimuli with a standard inter-stimulus interval (ISI) of 5 min. As in our previous study (Smolen et al. 2006), each individual tetanus is modeled as inducing: 1) A square-wave increase in synaptic $Ca^{2+}$ concentration, $[Ca_{syn}]$, to a peak value $A_{Casyn}$. The brief increase has a duration $Ca_{dur}$, with a standard value of 3 s that is similar to data (Malenka et al. 1992, Pologruto et al. 2004). Otherwise, $Ca^{2+}$ remains at a basal value $Ca_{bas}$; 2) An increase in cAMP concentration [cAMP] to a value $A_{cAMP}$, with duration $d_{cAMP}$. $d_{cAMP}$ has a standard value of 1 min, similar to data (Bacskai et al. 1993; Vincent and Brusciano 2001). Otherwise cAMP is at $cAMP_{bas}$; 3) An increase in the zeroth-order rate constant for activation of Raf kinase by its upstream effector Ras. This rate constant is denoted $k_{fRaf}$ and has a basal value of $k_{fbasRaf}$. Upon stimulus it increases to $(A_{STIM} + k_{fbasRaf})$. The duration of increase does not appear constrained by data. We assumed a plausible duration of 1 min.

CaMKII is assumed to be activated by a fourth-power Hill function of cytoplasmic $Ca^{2+}$ concentration. This Hill functions constitutes a minimal representation of the activation of CaM kinases by calmodulin (CaM), because four $Ca^{2+}$ ions bind cooperatively to CaM and CaM-$Ca_4$ activates CaM kinases. Data show a steep $[Ca^{2+}]$ dependence that can be characterized by a Hill coefficient ≥4 (Bradshaw et al. 2003).

$$f_1 = \frac{\left[Ca_{syn}\right]^4}{\left[Ca_{syn}\right]^4 + K_{Casyn}^4} \qquad 2)$$

The following ordinary differential equation (ODE) results for activation of CaMKII.

$$\frac{d\left[CaMKII_{act}\right]}{dt} = k_{fck2} f_1 - \frac{\left[CaMKII_{act}\right]}{\tau_{ck2}} \qquad 3)$$

For cAMP to activate PKA, two cAMP molecules must bind cooperatively to the regulatory subunit of a PKA holoenzyme (Herberg et al. 1996). Thus, our qualitative representation of PKA activation assumes the activation rate is a Hill function of the second power of [cAMP].

$$f_3 = \frac{[cAMP]^2}{[cAMP]^2 + K_{cAMP}^2} \tag{4}$$

$$\frac{d[PKA_{act}]}{dt} = \frac{(f_3 - [PKA_{act}])}{\tau_{PKA}} \tag{5}$$

In Eq. 4, the time constant $\tau_{PKA}$ is given a value of 15 min, resulting in substantial accumulation of PKA activity over three tetani with interstimulus intervals of 5 min.

At the synapse, Raf kinase is phosphorylated and activated by stimuli, and by a small basal rate constant.

$$[RAF] = RAF_{TOT} - [RAFP] \tag{6}$$

$$\frac{d[RAFP]}{dt} = k_{fRaf}[RAF] - k_{bRaf}[RAFP] \tag{7}$$

MAP kinase kinase, or MEK, is phosphorylated and activated by Raf. In turn, MEK activates ERK. Saturable Michaelis-Menten terms have been commonly used to describe phosphorylations and dephosphorylations of the kinases in the ERK cascade (e.g., Pettigrew et al. 2005), and are used here. Phosphatase dynamics are not explicitly modeled. Only doubly phosphorylated MEK and ERK are considered active. Concentrations of singly phosphorylated forms are given by conservation conditions.

$$\frac{d[MEK]}{dt} = -k_{fMEK}[RAFP]\frac{[MEK]}{[MEK] + K_{MEK}} + k_{bMEK}\frac{[MEKP]}{[MEKP] + K_{MEK}} \tag{8}$$

$$[MEKP] = MEK_{TOT} - [MEK] - [MEKPP] \tag{9}$$

$$\frac{d[MEKPP]}{dt} = k_{fMEK}[RAFP]\frac{[MEKP]}{[MEKP] + K_{MEK}} - k_{bMEK}\frac{[MEKPP]}{[MEKPP] + K_{MEK}} \tag{10}$$

$$\frac{d[ERK]}{dt} = -k_{fERK}[MEKPP]\frac{[ERK]}{[ERK] + K_{ERK}} + k_{bERK}\frac{[ERKP]}{[ERKP] + K_{ERK}} \tag{11}$$

$$[ERKP] = ERK_{TOT} - [ERK] - [ERKPP] \qquad (12)$$

$$\frac{d[ERKPP]}{dt} = k_{fERK}[MEKPP]\frac{[ERKP]}{[ERKP] + K_{ERK}} - k_{bERK}\frac{[ERKPP]}{[ERKPP] + K_{ERK}} \qquad (13)$$

As discussed previously (Smolen et al. 2006, 2014), data suggest that CaMKII, PKA, and ERK all play roles in setting the synaptic tag required for induction of late LTP. The molecular nature of the tag has not been characterized. We assume three phosphorylation sites, one for each kinase, and represent the amount of active synaptic tag as the product of the fractional phosphorylations of each site. The differential equations for the dynamics of these fractional phosphorylations, and their multiplication to give the level of active tag, are as in (Smolen et al. 2014).

$$\frac{d(Tag\text{-}1)}{dt} = k_{phos1}[CaMKII_{act}](1 - Tag\text{-}1) - k_{deph1}Tag\text{-}1 \qquad (14)$$

$$\frac{d(Tag\text{-}2)}{dt} = k_{phos2}[PKA_{act}](1 - Tag\text{-}2) - k_{deph2}Tag\text{-}2 \qquad (15)$$

$$\frac{d(Tag\text{-}3)}{dt} = k_{phos3}[ERKPP](1 - Tag\text{-}3) - k_{deph3}Tag\text{-}3 \qquad (16)$$

$$TAG = (Tag\text{-}1)(Tag\text{-}2)(Tag\text{-}3) \qquad (17)$$

PKMζ is translated after stimulus and is constitutively active, with activity denoted $PKM_{act}$. Its translation is increased by activation of CaMKII and ERK. As a minimal representation of these dynamics, CaMKII and ERK are each assumed to phosphorylate sites denoted $P_{CK2}$ and $P_{ERK}$. The rate of synthesis of active PKMζ is assumed proportional to the product of $P_{CK2}$ and $P_{ERK}$, along with a small basal rate of synthesis, $k_{transbaspkm}$.

$$\frac{d(P_{CK2})}{dt} = k_{phos4}[CaMKII_{act}](1 - P_{CK2}) - k_{deph4}P_{CK2} \qquad (18)$$

$$\frac{d(P_{ERK})}{dt} = k_{phos5}[ERKPP](1 - P_{ERK}) - k_{deph5}P_{ERK} \qquad (19)$$

$$\frac{d[PKM_{act}]}{dt} = k_{transpkm}P_{CK2}P_{ERK} + k_{transbaspkm} - k_{dpkm}[PKM_{act}] \qquad (20)$$

As in Smolen et al. (2006), the model hypothesizes a limiting factor, a protein $P_{lim}$, to prevent the synaptic weight W from increasing indefinitely with repeated simulation. As W increases, $P_{lim}$ is posited to decrease via incorporation into the synapse. The rate of increase of W is given as a product of the amount of TAG, the concentration of the plasticity-related protein PRP, the activity of PKMζ, and the concentration of the limiting protein $P_{lim}$. There is also a small basal rate of increase to balance ongoing decay.

$$\frac{dW}{dt} = k_{ltp} TAG [PRP][PKM_{act}] \frac{[P_{\lim}]}{[P_{\lim}]+K_{\lim}} + k_{ltpbas} - \frac{W}{\tau_{ltp}} \quad \text{21)}$$

The rate of $P_{\lim}$ consumption is assumed to increase as the rate of W increases, yielding a similar ODE. However, because PKMζ is assumed to increase W by enhancing AMPAR incorporation into the synapse (Migues et al. 2010), PKMζ is not included in the ODE for $P_{\lim}$ (i.e., $P_{\lim}$ is not taken to be AMPAR).

$$\frac{d[P_{\lim}]}{dt} = -k_{Pl} TAG [PRP] \frac{[P_{\lim}]}{[P_{\lim}]+K_{\lim}} + k_{Plbas} - \frac{W}{\tau_{Pl}} \quad \text{22)}$$

**Simulation of positive feedback**

Equations 2 – 22 as written describe the induction of late LTP (L-LTP) that decays over several hours, without any positive feedback to sustain L-LTP indefinitely. In simulations of persistent L-LTP, three types of feedback were implemented: 1) Sustained autoactivation of CaMKII, 2) Sustained autoactivation of PKMζ, or 3) Ongoing reactivation of a potentiated synapse, resulting in recurrent reinforcement of L-LTP. Only one feedback type was implemented at a time. These feedback types were respectively implemented as follows, by modifying the above equations:

Type 1: The right-hand side of Eq. 3, describing the rate of activation of CaMKII, was increased by adding a quadratic Hill function describing autoactivation,

$$f_{CaMKII} = \frac{k_{CaMKII} [CaMKII_{act}]^2}{[CaMKII_{act}]^2 + K_{CaMKII}^2} \quad \text{23)}$$

Type 2: The right-hand side of Eq. 20, describing the rate of activation of PKMζ, was increased by adding a quadratic Hill function describing autoactivation,

$$f_{PKM} = \frac{k_{PKM} [PKM_{act}]^2}{[PKM_{act}]^2 + K_{PKM}^2} \quad \text{24)}$$

Type 3: Eqs. 2-22 were unaltered. Instead, ongoing brief, spontaneous elevations of $[Ca_{syn}]$, [cAMP], and the rate constant $k_{fRaf}$ for Raf activation were added. Similarly to our previous study (Smolen, 2007), positive feedback was implemented by making the amplitudes of these elevations proportional to an increasing Hill function of synaptic weight. For this qualitative model, the Hill function was chosen to be fairly steep (power of 5) so that only potentiated synapses undergo significant reactivation. During each brief reactivation event, the elevated levels of $[Ca_{syn}]$, [cAMP], and $k_{fRaf}$ are,

$$Ca_{syn} = A_{Casyn} \frac{W^5}{W^5 + K_W^5}, \text{ with a floor of } Ca_{bas} \quad \text{25)}$$

$$[cAMP] = A_{cAMP} \frac{W^5}{W^5 + K_W^5}, \text{ with a floor of } cAMP_{bas} \quad \text{26)}$$

$$k_{fRaf} = k_{fbasRaf} + A_{STIM} \frac{W^5}{W^5 + K_W^5} \qquad 27)$$

The elevation duration was 3 s for Ca$_{syn}$ and 1 min for cAMP and k$_{fRaf}$. Outside of elevations [Ca$_{syn}$] = Ca$_{bas}$, [cAMP] = cAMP$_{bas}$, and k$_{fRaf}$ = k$_{fbasRaf}$.

In the absence of detailed data, the frequency of these reactivations was set low enough so that the synaptic tag variable TAG would relax to near its basal value between reactivations, to accord with current data that the tag decays within ~ 90 min after LTP induction (Frey and Morris 1997, 1998). The frequency was also set high enough to allow W to be maintained in an upper state for a reasonable value of the synaptic weight decay time constant $\tau_{ltp}$. To implement this, we used a Gaussian random variable for the intervals between successive reactivations. This variable was updated for each new reactivation, using the Box-Mueller algorithm. The mean interval was 500 min and the standard deviation was 150 min. A minimum interval of 10 min was imposed.

To simulate only induction, not maintenance, of L-LTP (Figs. 2A1-A2), Eqs. 23-27 were omitted.

**Parameter values and numerical methods**

Standard parameter values are used in all simulations except as noted otherwise. Without positive feedback, these are as follows:

Ca$_{bas}$ = 0.04 µM, cAMP$_{bas}$ = 0.06 µM, Ca$_{dur}$ = 3 s, d$_{cAMP}$ = 1 min, A$_{Casyn}$ = 0.8 µM, A$_{cAMP}$ = 0.25 µM, A$_{STIM}$ = 0.13 min$^{-1}$.

K$_{casyn}$ = 0.7 µM, k$_{fck2}$ = 180 µM min$^{-1}$, $\tau_{ck2}$ = 1 min, K$_{camp}$ = 1.0 µM, $\tau_{PKA}$ = 15 min.

RAF$_{TOT}$ = MEK$_{TOT}$ = ERK$_{TOT}$ = 0.25 µM, k$_{fbasRaf}$ = 0.0075 min$^{-1}$, k$_{bRaf}$ = 0.12 min$^{-1}$, k$_{fMEK}$ = 0.6 min$^{-1}$, k$_{bMEK}$ = 0.025 µM min$^{-1}$, K$_{MEK}$ = 0.25 µM, k$_{fERK}$ = 0.52 min$^{-1}$, k$_{bERK}$ = 0.025 µM min$^{-1}$, K$_{ERK}$ = 0.25 µM.

k$_{transpkm}$ = 0.2 µM min$^{-1}$, k$_{transbaspkm}$ = 0.0015 µM min$^{-1}$, k$_{dpkm}$ = 0.02 min$^{-1}$.

k$_{phos1}$ = 0.15 µM$^{-1}$ min$^{-1}$, k$_{deph1}$ = 0.008 min$^{-1}$, k$_{phos2}$ = 0.8 µM$^{-1}$ min$^{-1}$, k$_{deph2}$ = 0.2 min$^{-1}$, k$_{phos3}$ = 0.06 µM$^{-1}$ min$^{-1}$, k$_{deph3}$ = 0.05 min$^{-1}$, k$_{phos4}$ = 0.1 µM$^{-1}$ min$^{-1}$, k$_{deph4}$ = 0.1 min$^{-1}$, k$_{phos5}$ = 2.0 µM$^{-1}$ min$^{-1}$, k$_{deph5}$ = 0.1 min$^{-1}$.

[PRP] = 1.0 µM, k$_{ltp}$ = 500 µM$^{-2}$ min$^{-1}$, k$_{ltpbas}$ = 0.01 min$^{-1}$, $\tau_{ltp}$ = 300 min, K$_{lim}$ = 0.2 µM, k$_{Pl}$ = 6.0 min$^{-1}$, k$_{Plbas}$ = 0.0035 µM min$^{-1}$, $\tau_{Pl}$ = 100 min.

Positive feedback introduces new parameters (Eqs. 23-27). For positive feedback via CaMKII autoactivation, or via PKMζ autoactivation, new standard parameter values are, k$_{CaMKII}$ = 4.0 µM min$^{-1}$, K$_{CaMKII}$ = 1.0 µM, k$_{PKM}$ = 0.028 µM min$^{-1}$, K$_{PKM}$ = 0.75 µM. Also, k$_{ltp}$ = 70 µM$^{-2}$ min$^{-1}$ for CaMKII autoactivation, and 300 µM$^{-2}$ min$^{-1}$ for the simulation of Fig. 2B with PKMζ autoactivation. For the simulation of Fig. 2C with PKMζ increasing the synaptic tag, k$_{ltp}$ = 240 µM$^{-2}$ min$^{-1}$, and the coupling coefficient k$_{TPKM}$ = 0.0001 µM$^{-1}$.

For feedback via synaptic reactivation k$_{ltp}$ = 480 µM$^{-2}$ min$^{-1}$, K$_W$ = 4.0. $\tau_{ltp}$ was increased to 3200 min and k$_{ltpbas}$ was correspondingly decreased, to 0.00086 min$^{-1}$.

Model variant IV (Fig. 5) used the parameter values of the synaptic reactivation simulation (Fig. 3C) with the following exceptions. Reactivation events were weakened by reducing $A_{cAMP}$, $A_{Casyn}$, and $A_{STIM}$ (Eqs. 25-27) to 80% of their standard values. PKMζ positive feedback was included (Eq. 24), but the coupling constant $k_{PKM}$ was reduced to 0.0078 μM min$^{-1}$. These changes ensured neither form of positive feedback could sustain bistability on its own. Other changes were: $k_{transbaspkm}$ = 0.00045 μM min$^{-1}$, $k_{dpkm}$ = 0.006 min$^{-1}$, $k_{ltp}$ = 180 μM$^{-2}$ min$^{-1}$, $k_{ltpbas}$ = 0.003 min$^{-1}$, $τ_{ltp}$ = 800 min. The mean and standard deviation of reactivation frequency were (200 min, 50 min).

For some parameters, empirical data were used to determine standard values. As described in Smolen et al. (2006), the model's total concentrations of Raf, MEK, and ERK, and the peak concentration of active PKA as determined by stimulus and kinetic parameters, are similar to experimental values. The peak concentration of active CaMKII due to a tetanus is approximately 10% of the estimated total CaMKII concentration (Bhalla and Iyengar 1999). Data were also used to constrain the amplitude and duration of Ca$^{2+}$ and cAMP elevations (Smolen et al. 2006).

However, for many parameters, such as biochemical rate constants *in vivo*, data are lacking to determine values. Therefore, these parameters were chosen by trial and error to yield dynamics similar to empirical time courses, when the latter are available. In simulations that do not assume bistable CaMKII activity, CaMKII decays with a time constant of ~1 min. This brief lifetime is consistent with data describing CaMKII activity in dendritic spines (Murakoshi et al. 2017). PKA activity increases by ∼150% during L-LTP induction, and decays post-tetanus with a time constant of 15 min. This amplitude, and decay time constant, are consistent with data (Roberson and Sweatt 1996). Simulated postsynaptic ERK activation lasts ∼90 min post-tetanus. This duration is plausible but is bracketed by empirical data. One study suggests ERK remains phosphorylated for as long as 8 h after tetanus (Ahmed and Frey 2005). However, other studies (English and Sweatt 1997; Liu et al. 1999) suggest briefer activation of ∼30 min. Because long-lasting ERK activity could regulate transcription important for L-LTP, we suggest further study of ERK kinetics is warranted. After tetanus the synaptic tag (variable TAG) returns to baseline, or decays to a persistent but low activity to maintain L-LTP, within ~3 h. This decay time is similar to data (Frey and Morris 1997, 1998). L-LTP induction (the increase of W) is complete in ∼2 h, similar to the time required for empirical induction of L-LTP with BDNF (bypassing early LTP) (Ying et al. 2002). Parameters in the phenomenological representations of positive feedback loops (CaMKII autoactivation, persistent PKMζ activity, and synaptic reactivation) in Eqs. 23 – 27 are not determined by empirical data. Rather, values for these parameters were chosen by trial and error to yield bistable switches of synaptic weight that were induced by simulated tetani, but with both stable states robust to perturbations (weak stimuli). The resulting standard parameter value set cannot be claimed to be unique, and the simulated time courses of some variables, such as the individual TAG substrates Tag-1 – Tag-3, are not constrained by data.

In Figs. 2 - 5, some variables are vertically scaled, scale factors are as follows. Fig. 2A1-A2: 0.12 for [CaMKII$_{act}$], 7.0 for [ERKPP], 50.0 for [PKA$_{act}$], 0.1 for W, 700.0 for TAG, 2.0 for [PKM$_{act}$]. Fig. 2B1-B2, same as Fig. 2A1-A2 except 0.6 for [PKM$_{act}$]. Fig. 2C1-C2, same as Fig. 2B1-B2. Fig. 3A1-3A2, same as Fig. 2A1-A2 except 0.3 for [PKM$_{act}$]. Fig. 3B1-B2: 0.12 for [CaMKII$_{act}$], 11.0 for [ERKPP], 50.0 for [PKA$_{act}$], 0.1 for W, 1000.0 for TAG, 1.5 for [PKM$_{act}$]. Fig. 4A, same as Fig. 3A1-3A2. Fig. 4B, same as Fig. 3B1-3B2. Fig. 5, same as Fig. 3B except 0.75 for [PKM$_{act}$].

The forward Euler method was used to integrate ODEs and, for the simulations of Figs. 2A-B, we verified that using the fourth-order Runge-Kutta method did not yield significant differences. The time step was 10 ms. Model variables were initialized to 0.0001 (μM or arbitrary units) excepting variables given by conservation conditions. Prior to stimulus, model variables were equilibrated for at least two simulated days. The slowest variables, W and [PKM$_{act}$], were set to equilibrium basal values determined by other variables, except for Figs. 4A-B for which W and [PKM$_{act}$], were initialized in their upper state. Models were programmed in Java.

Java programs to reproduce the simulations in Figs. 2A, 2B, 3A and 3B have been deposited in the Github online database (https://github.com/psmolen0/Smolen_Byrne_LTP_Model), as files fig2ab3a.java and fig3b.java. These programs use model variants I-III, and will also be deposited in ModelDB.

## Acknowledgements

This study was supported by NIH grant NS102490. We thank H. Shouval and W. Sossin for comments on an earlier version of the manuscript.